\documentclass[pre,twocolumn]{revtex4}

\usepackage{epsfig}
\usepackage{bm,amsmath}
\usepackage{hyperref}

\newcommand{\gdot}{\dot{\gamma}}
\renewcommand{\d}{\mathrm{d}}

\begin{document}

\title{Understanding the approximations of mode-coupling theory for sheared steady states of colloids}
\author{Saroj Kumar Nandi}
\email{saroj@pks.mpg.de}
\affiliation{Max-Planck-Institute f{\"{u}}r Physik Komplexer Systeme, N{\"{o}}thnitzer Stra{\ss}e 38, 01187 Dresden, Germany}

\begin{abstract}
The lack of clarity of various mode-coupling theory (MCT) approximations, even in equilibrium, makes it hard to understand the relation between various MCT approaches for sheared steady states as well as their regime of validity. Here we try to understand these approximations indirectly by
deriving the MCT equations through two different approaches for a colloidal system under shear, first, through a microscopic approach, as suggested by Zaccarelli {\it et al}, and second, through fluctuating hydrodynamics, where the approximations used in the derivation are quite clear.
The qualitative similarity of our theory with a number of existing theories show that linear response theory might play a role in various approximations employed in deriving those theories and one needs to be careful while applying them for systems arbitrarily far away from equilibrium, such as a granular system or when shear is very strong.
As a byproduct of our calculation, we obtain the extension of Yvon-Born-Green (YBG) equation for a sheared system and under the assumption of random-phase approximation, the YBG equation yields the distorted structure factor that was earlier obtained through different approaches. 
\end{abstract}

\maketitle

\section{Introduction}

Shearing a supercooled fluid is ubiquitous in nature and has lots of technological applications \cite{larsonbook} for example industrial processing, testing usefulness of materials (e.g. paints or printing inks), 
mixing or separation of granular materials (in drug industry) etc. 
Shear starts affecting the properties of the system when $\tau_R\gdot\sim1$ where $\tau_R$ is the relaxation time of fluid and $\gdot$ is the rate of shear. As $\tau_R$ for glassy materials becomes
very large \cite{giuliormp}, even a small amount of shear will have large effect in the properties of the system. Shear can lead to interesting effects in a dense glassy system like
the shear induced crystallization and phase separation \cite{prachi11}, shear banding \cite{besseling10,aradian06}, shear thinning \cite{grest09,kobelev05,dullens11,sriram86,indrani95},
shear thickening \cite{krishnamurthy05,wagner09,kalman09,seto13} etc. But, understanding the properties of a system under shear is a nontrivial task as shear drives the system
out of equilibrium. Considerable progress has been achieved though in the past decade \cite{miyazaki02,miyazaki04,fuchs02,fuchs03,fuchs05}, mainly for colloidal glasses. 
Glass transition is defined as the point where the relaxation time of the system becomes of the order of $100s$. However, the relaxation time scale for molecular glasses far away from
the transition is $\sim 10^{-12}s$ and that for the colloids is $\sim 1ms$. For a consistent definition, the ratio of the relaxation times far away from the transition should be
comparable to that close to the transition. From that point of view, the glass transition for colloids should be defined when the relaxation time becomes $\sim 10^{11}s$. But this 
is practically impossible to measure. What this implies is that the colloidal glass is much further away from the point of its structural arrest compared to molecular glasses \cite{durian99}.
This raises the concern if a theory that has been successful for colloidal glasses can also be applied for other glassy systems. In any case, it is important to understand the 
approximations and assumptions made within a theory to infer its domain of applicability and how to extend it further. 

Mode-coupling theory (MCT) has been very successful in describing the supercooled and dense fluids \cite{goetzebook,das04,birolirmp,reichman05}. MCT gives an equation of motion
for the two-time density correlation function and makes several predictions that can be tested in experiments and simulations \cite{goetze85,das86,leutheusser84}. The correlation function
shows a complex two step relaxation near the glass transition point, first relaxing towards a plateaux, known as the $\beta$-relaxation, and then relaxation from the plateaux towards zero known as the $\alpha$ relaxation.
As the control parameters like temperature (or density) is decreased (or increased) the plateaux extends and the relaxation times increase. Below a certain temperature (or above a 
certain density) the correlation function ceases to decay to zero and this is known as the non-ergodicity transition of MCT. However, no such transition is observed in real systems or simulations
and predictions of MCT start to fail around this transition. It is argued that activated processes, not included within MCT, are responsible for avoiding such transition in real systems \cite{goetze87,sarika08}.
In a system under shear though there is no such transition even in the absence of any activated events as shear smears out the transition at a time scale $\sim \mathcal{O}(\gdot^{-1})$.
Thus, MCT might work better for a system under shear.

MCT has indeed been extended for systems under shear \cite{verberg97,fuchs02,fuchs03,brader08,miyazaki02,miyazaki04,chong09,suzuki13}. However, the approximations used in various theories
and their domain of applicability is not very clear. Even for bulk MCT, the approximations
used for the derivation of the theory is not yet well-understood \cite{reichman05,goetzebook,das04} and the role of fluctuation-dissipation relation (FDR) within the theory is quite nontrivial 
\cite{kuni05,kim07,basu07}. This issue becomes even more severe for the sheared steady state as the system is away from equilibrium and one needs to be careful that the 
FDR is not used explicitly or implicitly within the theory. It would be desirable to have a derivation of the theory where the approximations are clearer in order to understand its applicability and 
limitations. Here we take up this goal. The approach {\it a la} Zaccarelli {\it et al} \cite{zaccarelli01,zaclong} is particularly nice in this regard where the various approximations 
of the theory is quite transparent.

Starting from the Newton's equations of motion for individual particles of a colloidal suspension, we derive the MCT equations using the linear-response theory. An important
step in this derivation is the form of a trial function (Sec. \ref{trial_form}) that yields the Yvon-Born-Green (YBG) equation for the sheared fluid. We obtain the same form of YBG 
equation also through hydrodynamic approach using the approximation of local equilibrium, and thus justify the use of the particular form of the trial function.
For further insights, we also derive the MCT equations for a sheared fluid starting with the equations of fluctuating hydrodynamics.
Both approaches yield identical results which are qualitatively similar (at least, within the schematic approximation) to some of the existing theories \cite{miyazaki02,miyazaki04,fuchs05,brader07}. 
This should imply that the applicability of linear-response theory is assumed in some of the approximations in these set of theories even though it is not apparently clear in their derivation.

Thus, we can summarize the main achievements of the present work: (1) We have derived MCT for sheared steady state through the use of linear-response theory that can be justified for colloidal system under small shear. The qualitative similarity of the theory to many existing theories shows that linear response theory might play a role in various approximations employed within those theories.
(2) As a byproduct, we have 
obtained an extension of YBG equation for sheared colloids. Rest of the paper is organised as follows: starting from the microscopic equations of motion for the individual particles of a colloidal suspension under shear, we obtain the equation of motion for the coarse-grained density in Sec. \ref{cg_density}. We propose a trial form for sheared fluid in Sec. \ref{trialform} and obtain the modified YBG equation for a sheared fluid. We justify the use of the trial function in Sec. \ref{ybg_hydro} by comparing the YBG equation obtained through the use of the proposed trial function with that obtained through the hydrodynamic approach starting from the distribution functions. We obtain the mode-coupling equation in Sec. \ref{mct_zac} and present another derivation of sheared MCT equations through the hydrodynamic approach in Sec. \ref{mct_hydro}. We conclude the paper by discussing our results and their implications in Sec. \ref{discussion}.

\section{The equation of motion for the microscopic density}
\label{cg_density}
\begin{figure}[h]
\begin{center}
\includegraphics[height=4cm]{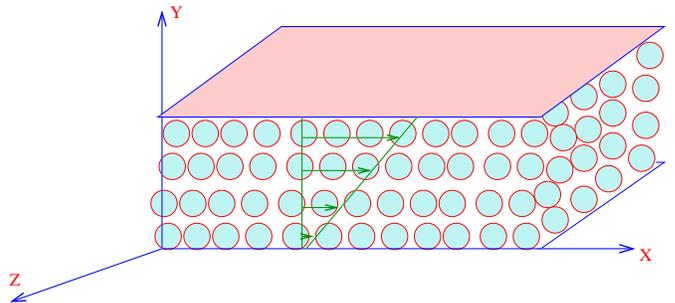}
\end{center}
\caption{The schematic picture of shearing a colloidal suspension taken within two plates. The upper plate is being sheared in $x$-direction at a rate $\dot{\gamma}$ and the velocity 
gradient is in $y$-direction.}
\label{shearschematic}
\end{figure}

Let us consider a three dimensional colloidal suspension between two plates and the upper plate is being sheared in the $x$-direction at a rate $\dot{\gamma}$ as schematically 
shown in Fig. \ref{shearschematic}.
The equations describing the $i$th particle of the fluid under steady shear in the frame of reference co-moving with the shear velocity are
\begin{align}
\dot{x}_i&=\dot{\gamma}y_i+P_{xi}/m; ~~~\dot{y}_i=P_{yi}/m; ~~~ \dot{z}_i=P_{zi}/m \nonumber\\
\dot{P}_{xi}&=F_{xi}-\dot{\gamma}P_{yi}-\zeta P_{xi}; ~\dot{P}_{yi}=F_{yi}-\zeta P_{yi}; \nonumber\\
\dot{P}_{zi}&=F_{zi}-\zeta P_{zi},
\end{align}
where $\dot{\gamma}y$ term in the $x$-component velocity equation of Eq. (1) comes from the contribution due to shear, $\zeta$ is a bare damping coefficient for the colloidal 
particles and $m$ is mass, same for all  particles. $\mathbf{P}_i=(P_{xi},P_{yi},P_{zi})$ is the momentum of $i$th particle and $\mathbf{F}_i$ is the inter-atomic force acting on it.

Let us now write down the equations of motion for the individual particles in the laboratory frame of reference as
\begin{align}
\dot{p}_{xi}&=\dot{P}_{xi}+\dot{\gamma}m\dot{y}_i =F_{xi}-\dot{\gamma}P_{yi}-\zeta P_{xi}+\dot{\gamma}P_{yi} \nonumber\\
&=F_{xi}-\zeta p_{xi}+\zeta m\dot{\gamma}y_i, \nonumber\\
\dot{p}_{yi}&=\dot{P}_{yi}=F_{yi}-\zeta p_{yi}; \hspace{0.4cm} \dot{p}_{zi}=\dot{P}_{zi}=F_{zi}-\zeta p_{zi},
\end{align}
where the $\mathbf{p}$'s are measured in the laboratory frame of reference and the $\mathbf{P}$'s are in the co-moving reference frame.

Then, in the vectorial form, the equation of motion for the $i$-th particle in the laboratory frame of reference can be written as
\begin{equation}
\label{force}
\dot{\bf p}={\bf F}-\zeta {\bf p}+\zeta m \dot{\gamma}y \hat{x}.
\end{equation}

The coarse-grained density in Fourier space at wave-vector $\mathbf{k}$ is
\begin{equation}
\rho_{\bf k}(t)=\sum_je^{i{\bf k}\cdot{\bf r}_j(t)}.
\end{equation}

Now for a system under shear we need to take into account the advection of wave vector. Due to shear, the system looses translational invariance, but it is restored by a 
Galilean transformation
\begin{equation}
{\bf k}(t)={\bf k}(0)+\dot{\gamma}tk_x\hat{y}
\end{equation}
for the kind of shear we are taking into account, namely, shear in $x$-direction and the velocity gradient in $y$-direction. For the convenience of notation, we will omit the 
time index for wave vectors below, since we are writing all the quantities at time $t$ only. The time derivative of the density in the co-moving reference frame will be
\begin{equation}
\dot{\rho}_{\bf k}(t)=\sum_j i({\bf k}\cdot\dot{\bf r}_j(t))e^{i{\bf k}\cdot{\bf r}_j(t)}
\end{equation}
and the second order time derivative will be
\begin{equation}
\label{secondder}
\ddot{\rho}_{\bf k}(t)=\sum_j i({\bf k}\cdot\ddot{\bf r}_j(t))e^{i{\bf k}\cdot{\bf r}_j(t)}-\sum_j({\bf k}\cdot\dot{\bf r}_j(t))^2e^{i{\bf k}\cdot{\bf r}_j(t)}.
\end{equation}
These equations are true for all wave vectors. Again for notational simplicity, we haven't time labeled the wave vectors, but an wave vector associated to a particular quantity is at the same time as the quantity is.

Now we will use Eq. (\ref{force}) to replace the $\ddot{\bf r}_j(t)$ term in the above equation and we will write down the equation in laboratory reference frame. The inter atomic potential is given by $U=\frac{1}{2}\sum_{m,m'}v(|{\bf r}_m(t)-{\bf r}_{m'}(t)|)=\frac{1}{2V}\sum_{m,m',{\bf k}'}v_{{\bf k}'}e^{-i{\bf k}'\cdot({\bf r}_m(t)-{\bf r}_{m'}(t))}$. Then the force on the $j$th particle is given as ${\bf F}_j=-\partial U/\partial {\bf r}_j(t)$ and following Ref. \cite{zaclong} we will obtain from Eq. (\ref{secondder})
\begin{eqnarray}
\label{order2derivative}
\ddot{\rho}_{\bf k}(t)&=&-\frac{1}{mV}\sum_{\bf k'}v_{\bf k'}{\bf k}\cdot{\bf k'}\rho_{{\bf k}-{\bf k'}}(t)\rho_{\bf k'}(t)-\zeta \frac{\partial {\rho}_{\bf k}(t)}{\partial t}\nonumber\\
&-&\sum_j ({\bf k}\cdot \dot{\bf r}_j(t))^2e^{i {\bf k}\cdot{\bf r}_j(t)}+\zeta \dot{\gamma}k_x \frac{\partial \rho_{\bf k}(t)}{\partial k_y}.
\end{eqnarray}
Here we have neglected higher order terms in $\gdot$.
This equation is exact but not of much use in its present form. To get useful insight about the dynamics of the system we need to write down an equation for the density dynamics 
separating the fast degrees of freedom from the slow ones and it is at this stage where various approximations enter. In the following, we will use similar approximations as 
used for the derivation of mode coupling theory for a bulk fluid without shear from the microscopic equations of motion \cite{zaccarelli01,zaclong}.


\section{The trial form and the Yvon-Born-Green equation}
\label{trial_form}
Let us first summarize the steps in the derivation of MCT for a {\em bulk unsheared} fluid following Zaccarelli et al's approach \cite{zaccarelli01,zaclong}. This will help comparing the derivation of sheared MCT with the unsheared one. Starting from the Newtonian equations of motion for the individual particles, we first write down the equation of motion for the coarse grained density. Next we use a trial form for the equation of motion for the coarse grained density as $\ddot{\rho}_k(t)+\hat\Omega_k\rho_k(t)=\hat{\mathcal{F}}_k(t)$, where $\hat\Omega_k$ is the frequency term having the dimension of the square of frequency and $\hat{\mathcal{F}}_k(t)$ is the residual force, that contains both the fast degrees of freedom and the slow ones. Minimisation of the residual force with respect to the frequency term gives the optimised value for the frequency term. Minimisation of $\hat{\mathcal{F}}_k(t)$ also implies an orthogonality condition that in turn gives the YBG (Yvon-Born-Green) equation. Finally we write down the residual force as the sum of a damping term and noise and use of the fluctuation-dissipation relation gives the form of the damping coefficient.

In the last section we have obtained the equation of motion for the coarse grained density starting from the equations of motion for individual particles of a sheared system. Next we need to use a trial form to write down the equation of motion for the coarse grained density in the desired form. 
Let us propose the following trial form for a sheared supercooled fluid 
\begin{equation}
\label{trialform}
\ddot{\rho}_{\bf k}(t)+\zeta \left(\frac{\partial}{\partial t}-\dot{\gamma}k_x \frac{\partial}{\partial k_y}\right)\rho_{\bf k}(t)+\hat{\Omega}_k \rho_{\bf k}(t)=\hat{\mathcal{F}}_{\bf k}(t),
\end{equation}
where $\zeta$ is the bare friction coefficient, $\dot{\gamma}$, the shear rate, $\hat{\Omega}_k$ has the dimension of square of frequency, $\hat{\mathcal{F}}_{\bf k}(t)$, the residual forces. In absence of the residual forces and shear, density waves would have shown a perfectly oscillatory behaviour, but shear damps the waves whereas the residual forces are responsible for deviation from an oscillatory behaviour of density waves.

This trial form for the case of sheared fluid is, of course, not obvious and we will justify the form by deriving the YBG equation from the above equation through the standard 
prescription, first suggested by Zwanzig \cite{zwanzig67}, and comparing that with the YBG equation derived from another completely independent approach, starting from 
distribution function \cite{hansenmcdonald} or the phase-space probability density. The YBG equation derived from distribution function requires the approximation of local equilibrium, which is justifiable only for small shear. We will see that the YBG equations derived from these two completely different approaches are the same and justifies the use of the above trial form for the case of sheared supercooled fluid.

Zwanzig calls the $\hat{\Omega}_k$'s as the elementary excitations of fluid\cite{zwanzig67} and suggests the variational principle to calculate the actual frequencies that are 
the eigenvalues of the Liouville operator. Thus we will minimize the residual force with respect to $\hat{\Omega}_k$ to get the value for the square of the frequency which 
should enter the actual equation of motion. The minimization of residual force\cite{zwanzig67,zaclong} implies
\begin{equation}
\frac{\partial \langle|\hat{\mathcal{F}}_{\bf k}(t)|^2\rangle}{\partial \hat\Omega_k}=0.
\end{equation}
Here all the averages are over the initial condition. Then, we will obtain the optimized frequency from the equation
\begin{align}
\langle\rho_{\bf k}(t)\ddot{\rho}_{-\bf k}(t)\rangle&+\zeta\langle\rho_{\bf k}(t)\bigg(\frac{\partial}{\partial t}-\dot{\gamma}k_x\frac{\partial}{\partial k_y}\bigg)\rho_{-\bf k}(t)\rangle \nonumber\\
&+\Omega_k\langle\rho_{\bf k}(t)\rho_{-\bf k}(t)\rangle=0,
\end{align}
and using the assumption that the fluid obeys the equipartition theorem at a temperature $T$, we obtain the frequency as
\begin{equation}
\label{frequency}
\Omega_{k}=\frac{k^2k_BT}{mS_{k}}+\frac{\zeta \dot{\gamma}k_x}{2S_{k}}\frac{\partial S_{k}}{\partial k_y}.
\end{equation}
This equation is true for all wave vectors and we have the definition of the distorted structure factor as $S_{k(t)}=\frac{1}{N}\langle\rho_{{\bf k}}(t)\rho_{-{\bf k}}(t)\rangle$. The distorted quantities are calculated from the input of the undistorted quantities and the theory gives an explicit way to calculate these quantities as we will see below.

Minimization of the residual force immediately gives an orthogonality condition between the residual force and the density as
\begin{equation}
\langle\rho_{-{\bf k}}(t)\mathcal{F}_{{\bf k}}(t)\rangle=0.
\end{equation}

Using Eq. (\ref{trialform}) we will have the orthogonality condition as
\begin{widetext}
\begin{align}
\Omega_k\langle\rho_{-{\bf k}}(t)\rho_{\bf k}(t)\rangle+\langle\rho_{-{\bf k}}(t)\ddot{\rho}_{\bf k}(t)\rangle +\zeta\langle\rho_{-{\bf k}}(t)\left(\frac{\partial}{\partial t}-\dot{\gamma}k_x\frac{\partial}{\partial k_y}\right)\rho_{\bf k}(t)\rangle=0.
\end{align}
After using the detailed form of $\ddot{\rho}_{\bf k}(t)$ derived above in Eq. (\ref{order2derivative}), we will get the equation as
\begin{align}
\Omega_k\langle\rho_{-{\bf k}}(t)\rho_{\bf k}(t)\rangle-\langle\sum_j({\bf k}\cdot{\dot{\bf r}_j(t)})^2e^{i{\bf k}\cdot{\bf r}_j(t)}\rho_{-{\bf k}}(t)\rangle 
-\frac{1}{mV}\sum_{{\bf k}'}v_{{\bf k}'}({\bf k}\cdot{\bf k}')\langle\rho_{-{\bf k}}(t)\rho_{{\bf k}-{\bf k}'}(t)\rho_{{\bf k}'}(t)\rangle=0  
\end{align}

While calculating averages like the second term in the above equation, we have explicitly assumed that the momenta and coordinate are uncorrelated. In general they are not 
as discussed by Cates and Ramaswamy in Ref. \cite{srmc}. But if we insist that they are uncorrelated, we will lose the long time hydrodynamic tail and in the limit of 
low inertia, which is true in the supercooled regime of the fluid that we are interested in, this assumption seems reasonable. 

Then we can write down the above expression as
\begin{align}
\Omega_kNS_k-\frac{k^2k_BT}{m}NS_k =\frac{1}{mV}\sum_{{\bf k}'}v_{{\bf k}'}({\bf k}\cdot{\bf k}')\langle\rho_{-{\bf k}}(t)\rho_{{\bf k}-{\bf k}'}(t)\rho_{{\bf k}'}(t)\rangle=0.
\end{align}
Using the expression for $\Omega_k$ from Eq. (\ref{frequency}) we will have
\begin{align}
\frac{k^2k_BT}{m}NS_k\bigg(\frac{1}{S_k}-1\bigg)+\frac{\zeta\dot{\gamma}k_xN}{2}\frac{\partial S_k}{\partial k_y}=\frac{1}{mV}k^2N^2S_kv_k 
+\frac{1}{mV}\sum_{{\bf k}'\neq{\bf k}}v_{{\bf k}'}({\bf k}\cdot{\bf k}')\langle\rho_{-{\bf k}}(t)\rho_{{\bf k}-{\bf k}'}(t)\rho_{{\bf k}'}(t)\rangle,
\end{align}
where in the above expression we have isolated the ${\bf k}'={\bf k}$ term from the sum and all the wave vectors are at time $t$. The structure factors $S_k$ in the above expression are the distorted ones and we have the relation between the direct correlation function $c_k$ and the structure factor as $S_k=1/(1-\rho c_k)$ where $\rho$ is the uniform density of the fluid. Using this relation, we will get the final expression as
\begin{align}
c_{{\bf k}}=-\beta v_{{\bf k}}+\frac{\beta m \zeta \dot{\gamma}k_x}{2k^2S_{k}\rho}\frac{\partial S_{k}}{\partial k_y}-\frac{\beta}{k^2N^2S_{k}}\sum_{{\bf k}'\neq{\bf k}}v_{{\bf k}'}({\bf k}\cdot{\bf k}')\langle \rho_{-{\bf k}}(t)\rho_{{\bf k}-{\bf k}'}(t)\rho_{{\bf k}'}(t)\rangle.
\end{align}
\end{widetext}
The above equation gives the relationship between the two point and three point correlation functions and expresses the coarse-grained macroscopic quantity, the direct correlation function, in terms of the microscopic quantity, the inter-atomic interaction potential.

The above equation is the YBG (Yvon-Born-Green) equation for a supercooled fluid under shear. The equation is modified from that of an unsheared fluid by the additional second term in the right hand side. If we use RPA (random phase approximation), the third term will drop out and we will be left with the simpler form of the equation as
\begin{equation}
\label{ybgwithrpa}
c_{{\bf k}}=-\beta v_{{\bf k}}+\frac{\beta m \zeta \dot{\gamma}k_x}{2k^2S_{k}\rho}\frac{\partial S_{k}}{\partial k_y}.
\end{equation}

This equation expresses how the distorted structural quantities, the direct correlation function $c_k$ and the structure factor $S_k$ of a sheared fluid are related to the 
microscopic interaction potential $v_k$. Now, the inter-atomic interaction potential $v_k$ does not get modified much due to shear. For a bulk unsheared fluid, we know that 
$c_k^{(0)}=-\beta v_k$, where $c_k^{(0)}$ is the direct correlation function under no shear. Therefore, using this equation in Eq. (\ref{ybgwithrpa}) we can obtain the 
information of the distorted structure factor from $S_k^{(0)}$, the undistorted one. After a formal manipulation of the equation we will obtain the distorted structure factor as
\begin{equation}
\label{shearsk}
S_k=S_k^{(0)}+S_k^{(0)}\frac{\beta m\zeta \dot{\gamma}k_x}{2k^2}\frac{\partial S_k}{\partial k_y}.
\end{equation}
To solve the mode coupling equation we need the information of distorted structure factor as input and the above equation gives us this quantity in terms of the undistorted ones. The same expression was obtained earlier for colloidal suspensions \cite{indrani95,ronis84}.


\section{YBG equation starting from distribution function}
\label{ybg_hydro}
As we have discussed above, the use of the particular trial form for the coarse-grained density equation of motion is not obvious. We will justify this particular trial form by comparing the YBG equation derived above with that derived from a completely different approach, starting from the distribution function \cite{hansenmcdonald}.
In the second approach, we don't need any other assumptions apart from that of the local equilibrium.

We have the distribution function or phase-space probability density $f^{[N]}({\bf r}^N,{\bf p}^N;t)$, which gives the probability density that at time $t$, the physical system is found around a point $({\bf r}^N,{\bf p}^N)$ in the $6N$ dimensional phase space. Then, we must have, for all time $t$,
\begin{equation}
\int f^{[N]}({\bf r}^N,{\bf p}^N;t) d{\bf r}^Nd{\bf p}^N=1.
\end{equation}
The Liouville equation can be written as
\begin{equation}
\frac{\partial f^{[N]}}{\partial t}+\sum_{i=1}^N \left(\frac{\partial f^{[N]}}{\partial {\bf r}_i}\cdot \dot{\bf r}_i+\frac{\partial f^{[N]}}{\partial {\bf p}_i}\cdot \dot{\bf p}_i\right)=0,
\end{equation}
or, more compactly
\begin{equation}
\frac{\partial f^{[N]}}{\partial t}=\{\mathcal{H},f^{[N]}\},
\end{equation}
where $\{A,B\}$ denotes the Poisson bracket:
\begin{equation}
\{A,B\}\equiv \sum_{i=1}^N\left(\frac{\partial A}{\partial {\bf r}_i}\cdot \frac{\partial B}{\partial {\bf p}_i}-\frac{\partial A}{\partial {\bf p}_i}\cdot \frac{\partial B}{\partial {\bf r}_i}\right).
\end{equation}

The reduced phase-space distribution function for the $n$ particles, integrating out the position and momenta of the rest of the $(N-n)$ particles, is defined as
\begin{equation}
f^{(n)}({\bf r}^n,{\bf p}^n;t)=\frac{N!}{(N-n)!}\int f^{[N]}({\bf r}^N,{\bf p}^N;t)d{\bf r}^{(N-n)}d{\bf p}^{(N-n)}.
\end{equation}

In the laboratory frame of reference, the equation of motion of the $i$-th particle of the colloidal suspension under shear will be
\begin{equation}
\dot{\bf p}_i={\bf F}_i-\zeta{\bf p}_i+\zeta m \dot{\gamma}y_i\hat{x}
\end{equation}
for the particular kind of shearing shown in Fig. 1.

Then the $N$-particle distribution function will follow an equation given as
\begin{widetext}
\begin{equation}
\left[ \frac{\partial}{\partial t}+ \sum_{i=1}^N\frac{{\bf p}_i-\dot{\gamma}my_i\hat{x}}{m}\cdot\frac{\partial}{\partial {\bf r}_i}-\sum_{i=1}^N(\zeta {\bf p}_i-\zeta m \dot{\gamma}y_i\hat{x})\cdot\frac{\partial}{\partial {\bf p}_i} \right]f^{[N]}=-\sum_{i=1}^N\sum_{j=1}^N{\bf F}_{ij}\cdot\frac{\partial f^{[N]}}{\partial {\bf p}_i}.
\end{equation}
where all the quantities are written in the laboratory frame of reference.
Now, we multiply the above by $N!/(N-n)!$ and integrate over the $3(N-n)$ coordinates and $3(N-n)$ momenta. Then we will get
\begin{align}
\bigg(\frac{\partial }{\partial t}+\sum_{i=1}^n\frac{{\bf p}_i-\dot{\gamma}my_i\hat{x}}{m}\cdot\frac{\partial }{\partial {\bf r}_i}&+\sum_{i=1}^n \zeta({\bf p}_i-\dot{\gamma}my_i\hat{x}) \cdot \frac{\partial }{\partial {\bf p}_i}\bigg)f^{(n)}=-\sum_{i=1}^n\sum_{j=1}^n{\bf F}_{ij}\cdot \frac{\partial f^{(n)}}{\partial {\bf p}_i}\nonumber\\
&-\frac{N!}{(N-n)!}\sum_{i=1}^n\sum_{j=n+1}^N\int{\bf F}_{ij}\cdot \frac{\partial f^{[N]}}{\partial {\bf p}_i} d{\bf r}^{(N-n)}d{\bf p}^{(N-n)}.
\end{align}

We assume that the fluid is in a steady state and locally in equilibrium and. The first and the third terms in left hand side will become zero in steady state. The way to see why the third term is zero is as follows. Let us take the Fourier transform of the above equation and then the ${\bf p}f_k^{(n)}$ kind of term can be written as $\partial/\partial t (f_k^{(n)}e^{-i{\bf k}\cdot{\bf r}})$ and therefore, in the steady state, the time derivatives will have to be zero. We concentrate on the $n=2$ term to get the YBG equation. The last term in right hand side can be taken as the sum of $(N-n)$ identical terms and we can write the above equation as
\begin{align}
\label{distfunc}
\bigg(\frac{{\bf p}_1-\dot{\gamma}my_1\hat{x}}{m}\cdot\nabla_1&+\zeta\dot{\gamma}my_1\hat{x}\cdot\frac{\partial}{\partial {\bf p}_1}+{\bf F}_{12}\cdot\frac{\partial }{\partial {\bf p}_1}+\frac{{\bf p}_2-\dot{\gamma}my_2\hat{x}}{m}\cdot\nabla_2+\zeta\dot{\gamma}my_2\hat{x}\cdot\frac{\partial}{\partial {\bf p}_2} +{\bf F}_{21}\cdot\frac{\partial }{\partial {\bf p}_2}\bigg)f_0^{(2)} \nonumber\\
&=-\int{\bf F}_{13}\cdot\frac{\partial f_0^{(3)}}{\partial {\bf p}_1}d{\bf r}_3d{\bf p}_3-\int{\bf F}_{23}\cdot\frac{\partial f_0^{(3)}}{\partial {\bf p}_2}d{\bf r}_3d{\bf p}_3.
\end{align}

Now, at local equilibrium, we will have
\begin{eqnarray}
\label{mbdist}
f_0^{(2)}({\bf r}_1,{\bf r}_2,{\bf p}_1,{\bf p}_2)&=&\rho^{(2)}({\bf r}_1,{\bf r}_2)f_M({\bf p}_1)f_M({\bf p}_2)   \nonumber\\
f_0^{(3)}({\bf r}_1,{\bf r}_2,{\bf r}_3,{\bf p}_1,{\bf p}_2,{\bf p}_3)&=&\rho^{(3)}({\bf r}_1,{\bf r}_2,{\bf r}_3)f_M({\bf p}_1)f_M({\bf p}_2)f_M({\bf p}_3).
\end{eqnarray}
where $f_M({\bf p})$ is the Maxwell-Boltzmann distribution function and $\rho^{(n)}({\bf r}^n)$ is the $n$-particle density. Under shear, because of the advected velocity field, the Maxwell-Boltzmann velocity distribution function will be modified as
\begin{equation}
\label{fm}
f_M({\bf p}_i)=\frac{1}{(2\pi mk_BT)^{3/2}}e^{-\frac{\beta}{2m}({\bf p}_i-\dot{\gamma}my_i\hat{x})^2}
\end{equation}
and therefore we will have
\begin{equation}
\label{gradpfm}
\frac{\partial f_M({\bf p}_i)}{\partial {\bf p}_i}=-\frac{\beta}{m}({\bf p}_i-\dot{\gamma}my_i\hat{x})f_M({\bf p}_i).
\end{equation}

Using Eqs. (\ref{mbdist})-(\ref{gradpfm}) in Eq. (\ref{distfunc}), we will obtain
\begin{align}
\frac{({\bf p}_1-\dot{\gamma}my_1\hat{x})}{m}\cdot\bigg[(\nabla_1-&\beta\zeta\dot{\gamma}my_1\hat{x}-\beta{\bf F}_{12})\rho^{(2)}({\bf r}_1,{\bf r}_2)-\beta\int{\bf F}_{13}\rho^{(3)}({\bf r}_1,{\bf r}_2,{\bf r}_3)d{\bf r}_3\bigg] \nonumber\\
+\frac{({\bf p}_2-\dot{\gamma}my_2\hat{x})}{m}&\cdot\bigg[(\nabla_2-\beta\zeta\dot{\gamma}my_2-\beta{\bf F}_{21})\rho^{(2)}({\bf r}_1,{\bf r}_2) -\beta\int{\bf F}_{23}\rho^{(3)}({\bf r}_1,{\bf r}_2,{\bf r}_3)d{\bf r}_3\bigg]=0.
\end{align}
This equation can be thought of as the dot product of two 2$d$-dimensional vectors: ${\bf P}_i\cdot {\bf Q}=0$. Since this equation is true for any ${\bf P}_i$, we must have ${\bf Q}=0$. Then we have
\begin{align}
\label{ybgshear1}
(\nabla_1-\frac{\beta}{m}\zeta\dot{\gamma}my_1\hat{x}-\frac{\beta}{m}{\bf F}_{12})\rho^{(2)}({\bf r}_1,{\bf r}_2)-\frac{\beta}{m}\int{\bf F}_{13}\rho^{(3)}({\bf r}_1,{\bf r}_2,{\bf r}_3)d{\bf r}_3=0.
\end{align}

From the definitions of the $l$-particle distribution function, $g^{(l)}({\bf r}^l)$, we have
\begin{equation}
\rho^{(l)}({\bf r}^l)=\rho^lg^{(l)}({\bf r}^l)
\end{equation}
and the force is given as ${\bf F}_{ij}=-\nabla_iv({\bf r}_i,{\bf r}_j)$ and using these we will have from Eq. (\ref{ybgshear1}),
\begin{equation}
(k_BT\nabla_1-\zeta\dot{\gamma}my_1\hat{x}+\nabla_1 v({\bf r}_1,{\bf r}_2))g^{(2)}({\bf r}_1,{\bf r}_2)=-\rho\int\nabla_1 v({\bf r}_1,{\bf r}_3)g^{(3)}({\bf r}_1,{\bf r}_2,{\bf r}_3)d{\bf r}_3.
\end{equation}

 Next we take a dot product of the resulting equation with $\nabla_1$ and upon Fourier transforming we obtain
\begin{align}
-k_BTk^2g_k^{(2)}+\zeta m\dot{\gamma}k_x\frac{\partial g_k^{(2)}}{2\partial k_y}=\frac{1}{\rho^2V^2}\sum_{{\bf k}'}v_{{\bf k}'}({\bf k}\cdot{\bf k}')\langle \rho_{-{\bf k}}(t)\rho_{{\bf k}-{\bf k}'}(t)\rho_{{\bf k}'}(t)\rangle
\end{align}
where we have used the fact that $\rho^{(3)}({\bf k},{\bf k}')+\rho^{(2)}({\bf k})=\langle \rho_{-{\bf k}}(t)\rho_{{\bf k}-{\bf k}'}(t)\rho_{{\bf k}'}(t)\rangle$. Now we use the relation $g_k^{(2)}=(S_k-1)/\rho$ and write down the above equation as
\begin{align}
\label{ybg38}
-\frac{k^2S_k}{\rho\beta}\bigg(1-\frac{1}{S_k}\bigg)=-\frac{\zeta m\dot{\gamma}k_x}{2\rho}\frac{\partial S_k}{\partial k_y}+\frac{1}{N^2}\sum_{{\bf k}'}v_{{\bf k}'}({\bf k}\cdot{\bf k}')\langle \rho_{-{\bf k}}(t)\rho_{{\bf k}-{\bf k}'}(t)\rho_{{\bf k}'}(t)\rangle.
\end{align}
Using the relation $\rho c_k=1-1/S_k$ between the structure factor and the direct correlation function $c_k$, we obtain from Eq. (\ref{ybg38})
\begin{align}
c_{{\bf k}}=-\beta v_{{\bf k}}+\frac{\beta m \zeta \dot{\gamma}k_x}{2k^2S_{k}n}\frac{\partial S_{k}}{\partial k_y}-\frac{\beta}{k^2N^2S_{k}}\sum_{{\bf k}'\neq{\bf k}}v_{{\bf k}'}({\bf k}\cdot{\bf k}')\langle \rho_{-{\bf k}}(t)\rho_{{\bf k}-{\bf k}'}(t)\rho_{{\bf k}'}(t)\rangle
\end{align}
\end{widetext}
where we have separated out the ${\bf k}'={\bf k}$ term in the sum. This is the YBG equation as we have obtained earlier through a completely different approach using the trial form for the coarse grained density equation of motion. 
The assumptions used in the above derivation are that the distribution functions for coordinate and momenta factor out and that the velocity distribution function is governed by a Maxwell-Boltzmann distribution function with the mean being shifted to that of the imposed preferred velocity should hold in a steady state only if shear is not too high.
The fact that the YBG equation derived through the use of the proposed trial form is exactly same as the one derived through the standard route starting from distribution function justifies the particular form of the trial function used above for the coarse grained density equation of motion.

\section{The mode coupling equation for the sheared fluid}
\label{mct_zac}
Once we accept the trial form as in Eq. (\ref{trialform}), obtaining the mode coupling equation is fairly straightforward. But as we discussed earlier, one conceptual difficulty is the validity of fluctuation-dissipation relation (FDR). When in equilibrium, the noise is related to the dissipation coefficient through the FDR. Near the transition point, the structural relaxation time, $\tau$, of the fluid is quite high. Shear pumps energy into the system at a time scale $\dot{\gamma}^{-1}$ and this energy spreads in the system through the fast degrees of freedom. If the fluid is away from the transition point, one can safely assume FDR since the fast degrees of freedom are too fast to be affected by a small shear rate. For colloidal glasses, if the shear is not too high, one can still assume the validity of linear-response theory.

First we will divide the residual force in two parts: the frictional memory kernel and the noise. These two quantities are related by FDR. Thus, we have
\begin{widetext}
\begin{equation}
\label{zacmemory}
\mathcal{F}_{\bf k}(t)=-\int_0^t \gamma_{\bf k}(t-t')\dot{\rho}_{\bf k}(t')dt'+ f_{\bf k}(t), \,\, \text{ and }\gamma_{\bf k}(t)=\frac{\langle f_{{\bf k}(t)}(t)f_{-{\bf k}}(0)\rangle}{\langle|\dot{\rho}_{{\bf k}(t)}|^2\rangle}
\end{equation}
The explicit form of the noise term will be
\begin{align}
f_{{\bf k}(t)}(t)=\Omega_{{\bf k}(t)}&\rho_{{\bf k}(t)}(t)-\frac{1}{mV}\sum_{{\bf k}'}v_{{\bf k}'}({\bf k}(t)\cdot{\bf k}')\rho_{{\bf k}(t)-{\bf k}'}(t)\rho_{{\bf k}'}(t)\nonumber\\
&-\sum_j({\bf k}(t)\cdot\dot{\bf r}_j(t))^2e^{i{\bf k}(t)\cdot{\bf r}_j(t)}+\int_0^t\gamma_{{\bf k}(t)}(t-t')\dot{\rho}_{{\bf k}(t)}(t')dt'.
\end{align}

Here in the noise term we don't have the term $\zeta\left(\frac{\partial}{\partial t}-\dot{\gamma}k_x\frac{\partial}{\partial k_y}\right)\rho_{{\bf k}(t)}(t)$ because of the particular trial form we have opted. This term will get cancelled in the trial form with that coming from $\ddot{\rho}_{\bf k}(t)$ when we write the later in its detailed microscopic form.

In the two-time correlation functions, because of the advection of wave vectors, the wave vector ${\bf k}$ at time $t=0$ gets contribution from the wave vector ${\bf k}(t)$ at time $t$. Therefore, the dynamic structure factor is defined as
\begin{equation}
S_k(t)=\frac{1}{N}\langle \rho_{{\bf k}(t)}(t)\rho_{-\bf k}(0)\rangle.
\end{equation}
Using Eq. (\ref{zacmemory}), we will have the expression for memory kernel as
\begin{align}
\gamma_{\bf k}(t)&=\frac{\beta m}{k(t)^2N}\bigg[\Omega_{{\bf k}(t)}\Omega_{\bf k}\langle\rho_{{\bf k}(t)}(t)\rho_{-{\bf k}}(0)\rangle-\frac{\Omega_{\bf k}}{mV}\sum_{{\bf k}'}v_{{\bf k}'}({\bf k}(t)\cdot{\bf k}')\langle\rho_{{\bf k}(t)-{\bf }'}(t)\rho_{{\bf k}'}(t)\rho_{-{\bf k}}(0)\rangle \nonumber\\
-&\Omega_{\bf k}\langle\sum_j({\bf k}(t)\cdot\dot{\bf r}_j(t))^2e^{i{\bf k}(t)\cdot{\bf r}_j(t)}\rho_{-{\bf k}}(0)\rangle+\frac{\Omega_{{\bf k}(t)}}{mV}\sum_{{\bf k}''}v_{{\bf k}''}({\bf k}\cdot{\bf k}'')\langle\rho_{{\bf k}(t)}(t)\rho_{-{\bf k}-{\bf k}''}(0)\rho_{-{\bf k}''}(0)\rangle \nonumber\\
-&\frac{1}{(mV)^2}\sum_{{\bf k}',{\bf k}''}v_{{\bf k}'}v_{{\bf k}''}({\bf k}(t)\cdot{\bf k}')({\bf k}\cdot{\bf k}'')\langle\rho_{{\bf k}(t)-{\bf k}'}(t)\rho_{{\bf k}'}(t)\rho_{-{\bf k}-{\bf k}''}(0)\rho_{{\bf k}''}(0)\rangle \nonumber\\ 
-&\frac{1}{mV}\sum_{{\bf k}',j}v_{{\bf k}'}({\bf k}\cdot{\bf k}')\langle({\bf k}(t)\cdot\dot{\bf r}_j(t))^2e^{i{\bf k}(t).{\bf r}_j(t)}\rho_{-{\bf k}-{\bf k}'}(0)\rho_{{\bf k}'}(0)\rangle -\Omega_{{\bf k}(t)}\langle\sum_l({\bf k}\cdot\dot{\bf r}_l(0))^2\rho_{{\bf k}(t)}(t)e^{-i{\bf k}\cdot{\bf r}_l(0)}\rangle\nonumber\\
+&\frac{1}{mV}\sum_{{\bf k}',l}v_{{\bf k}'}({\bf k}(t)\cdot{\bf k}')\langle({\bf k}\cdot\dot{\bf r}_l(0))^2\rho_{{\bf k}(t)-{\bf k}'}(t)\rho_{{\bf k}'}(t)e^{-i{\bf k}\cdot{\bf r}_l(0)}\rangle 
+\langle\sum_{j,l}({\bf k}(t)\cdot\dot{\bf r}_j(t))^2({\bf k}\cdot\dot{\bf r}_l(0))^2e^{i{\bf k}(t)\cdot{\bf r}_j(t)}e^{-i{\bf k}\cdot{\bf r}_l(0)}\rangle \nonumber\\
+&\int_0^t\gamma_{{\bf k}(t)}(t-t')\bigg\langle\dot{\rho}_{{\bf k}(t')}(t')\big(\Omega_k\rho_{-{\bf k}}(0) +\frac{1}{mV}\sum_{{\bf k}'}v_{{\bf k}'}({\bf k}\cdot{\bf k}')\rho_{-{\bf k}-{\bf k}'}(0)\rho_{{\bf k}'}(0)
-\sum_l({\bf k}\cdot\dot{\bf r}_l(0))^2e^{-i{\bf k}\cdot{\bf r}_l(0)}\big)\bigg\rangle\bigg]
\end{align}

Calculating the various contributions from the above terms is quite straight forward, although a bit cumbersome. Let us concentrate on the last three terms. The first of these terms can be written as
\begin{equation}
\Omega_k\langle\dot{\rho}_{{\bf k}(t')}(t')\rho_{-{\bf k}}(0)\rangle=\Omega_kN\dot{S}_k(t').
\end{equation}
The penultimate term has a three point density which will be calculated as
\begin{eqnarray}
\langle\dot{\rho}_{{\bf k}(t')}(t')\rho_{-{\bf k}-{\bf k}'}(0)\rho_{{\bf k}'}(0)\rangle&=&\langle\dot{\rho}_{{\bf k}(t')}(t')\rho_{-{\bf k}-{\bf k}'}(0)\rangle\langle\rho_{{\bf k}'}(0)\rangle +\langle\dot{\rho}_{{\bf k}(t')}(t')\rho_{{\bf k}'}(0)\rangle\langle\rho_{-{\bf k}-{\bf k}'}(0)\rangle\nonumber\\
&=&N^2\dot{S}_k(t')\delta_{{\bf k}',0}+N^2\dot{S}_k(t')\delta_{{\bf k}',-{\bf k}}.
\end{eqnarray}
The first term doesn't contribute anything because the delta function kills the term through the factor sitting in front of this three point density correlator and the second term amounts to $-\frac{n}{m}v_{\bf k}k^2N\dot{S}_{k}(t')$.

The last term is written as 
\begin{equation}
-\langle\sum_l({\bf k}\cdot\dot{\bf r}_l(0))^2\dot{\rho}_{{\bf k}(t')}(t')e^{-i{\bf k}\cdot{\bf r}_l(0)}\rangle=-\sum_l\langle({\bf k}\cdot\dot{\bf r}_l(0))^2\rangle\langle\dot{\rho}_{{\bf k}(t')}(t')e^{-i{\bf k}\cdot{\bf r}_l(0)}\rangle=-\frac{k^2}{\beta m}N\dot{S}_{k}(t').
\end{equation}
These three terms, using the explicit forms of $\Omega_k$ and $v_{\bf k}$ with the approximation of RPA, Eq. (\ref{ybgwithrpa}), adds up to zero. Following similar manipulations we will obtain the memory kernel as
\begin{align}
\label{shearmemory1}
\gamma_{{\bf k}}(t)&=\frac{n\beta}{mk(t)^2V}\sum_{{\bf k}'(t)\neq{\bf k}(t)}\bigg[v_{{\bf k}'(t)}v_{{\bf k}'(0)}({\bf k}(t)\cdot{\bf k}'(t))({\bf k}(0)\cdot{\bf k}'(0)) \nonumber\\
&+v_{{\bf k}'(t)}v_{{\bf k}(0)-{\bf k}'(0)}({\bf k}(t)\cdot{\bf k}'(t))({\bf k}(0)\cdot({\bf k}(0)-{\bf k}'(0)))\bigg]S_{{\bf k}-{\bf k}'}(t)S_{{\bf k}'}(t),
\end{align}
where $v_{\bf k}$'s are to be replaced by the undistorted direct correlation function using the YBG equation.

Now with a transformation of variable and symmetrizing the terms, we can write down Eq. (\ref{shearmemory1}) as
\begin{align}
\label{secondform}
\gamma_{{\bf k}}(t) &=\frac{n\beta}{2mk(t)^2V}\sum_{{\bf k}'(t)}\bigg[v_{{\bf k}'}({\bf k}\cdot{\bf k}')+v_{{\bf k}-{\bf k}'}{\bf k}\cdot({\bf k}-{\bf k}')\bigg]\times \nonumber \\
&\bigg[v_{{\bf k}'(t)}({\bf k}(t)\cdot{\bf k}'(t))+v_{{\bf k}(t)-{\bf k}'(t)}{\bf k}(t)\cdot({\bf k}(t)-{\bf k}'(t))\bigg]S_{{\bf k}-{\bf k}'}(t)S_{{\bf k}'}(t)
\end{align}
where the wave vectors without any time indices are supposed to be at time $t=0$. Therefore the mode coupling equation will become
\begin{equation}
\label{shearMCT1}
\ddot{\phi}_{{\bf k}}(t)+\zeta\left(\frac{\partial }{\partial t}-\dot{\gamma}k_x\frac{\partial }{\partial k_y}\right)\phi_{{\bf k}}(t)+\Omega_{{\bf k}}\phi_{{\bf k}}(t)+\int_0^t\gamma_{{\bf k}}(t-t')\dot{\phi}_{{\bf k}}(t')dt'=0
\end{equation}
where $\phi_{{\bf k}}(t)=\langle\rho_{{\bf k}(t)}(t)\rho_{-{\bf k}}(0)\rangle/S_{{\bf k}}$.

The explicit form of the memory kernel is given by Eq. (\ref{secondform}). Now, as we have discussed earlier, the inter-atomic interaction potential is not affected much by the shear and therefore, we will replace $-\beta v_k$ by $c_k^{(0)}$, the undistorted direct correlation function which is an equilibrium relation under no shear. Then we will have, after replacing the sum by an integral,
\begin{align}
\label{shearMCT2}
\gamma_{\bf k}=&\frac{k_BT\rho_0}{2 k(t)^2}\int_{{\bf k}'}\bigg[c_{{\bf k}'}^{(0)}({\bf k}\cdot{\bf k}')+c_{{\bf k}-{\bf k}'}^{(0)}{\bf k}\cdot({\bf k}-{\bf k}')\bigg]\times \nonumber \\
&\bigg[c_{{\bf k}'(t)}^{(0)}({\bf k}(t)\cdot{\bf k}'(t))+c_{{\bf k}(t)-{\bf k}'(t)}^{(0)}{\bf k}(t)\cdot({\bf k}(t)-{\bf k}'(t))\bigg]S_{{\bf k}-{\bf k}'}(t)S_{{\bf k}'}(t).
\end{align}
Eq. (\ref{shearMCT1}) along with Eq. (\ref{shearMCT2}) constitutes the final MCT equations for a sheared fluid. Solving these equations requires the distorted static structure factor of a sheared fluid as input.

\section{Derivation of MCT equation through the hydrodynamic approach}
\label{mct_hydro}
To have further insight in to the theory, we obtain the sheared MCT through another approach, the fluctuating hydrodynamics.
The equations of fluctuating hydrodynamics for an isothermal fluid are the continuity equations for number density, $\rho({\bf x},t)$, and momentum density ${\bf g}({\bf x},t)$ at position $\mathbf{x}$ and time $t$:
\begin{subequations}
\begin{align}
\frac{\partial \rho({\bf x},t)}{\partial t} &= -\nabla \cdot (\rho({\bf x},t){\bf v}({\bf x},t)) \label{densityeq}\\
\frac{\partial {\bf g}({\bf x},t)}{\partial t}+\nabla\cdot({\bf g}{\bf v}({\bf x},t))&=-\nabla p({\bf x},t)+\eta\bigtriangledown^2 \mathbf{v}+(\zeta+\eta/3)\nabla(\nabla\cdot{\bf v})+{\bf f}({\bf x},t)
\label{momentumeq}
\end{align}
\end{subequations}
where ${\bf g}({\bf x},t)=\rho({\bf x},t){\bf v}({\bf x},t)$, $p$ is the pressure, and $\eta$ and $\zeta$ are shear and bulk viscosities respectively. We set the particle mass to 
unity and therefore the mass density and number density are same. The noise must satisfy
\begin{equation}
\langle f_i({\bf x},t)f_j({\bf x}',t')\rangle=2k_BT(\eta\bigtriangledown^2 \delta_{ij}+(\zeta+\eta/3)\nabla_i\nabla_j)\delta({\bf x}-{\bf x}')\delta(t-t'),
\end{equation}
where $k_BT$ is Boltzmann's constant times the temperature.
The pressure term in the momentum equation comes as a pure gradient which is sufficient when we look at a length scale much larger than the individual molecular diameter. However, if we look at a phenomenon occurring at the molecular length scale, as the glass transition is, we must replace this term by the local force density which is the local density times the gradient of the local chemical potential, $\rho({\bf x},t)\nabla\mu$. The functional derivative of a suitably chosen free energy $F[\rho]$ with respect to the local density is the local chemical potential and therefore the force density becomes $\rho({\bf x},t)\nabla\frac{\delta F[\rho]}{\delta \rho({\bf x},t)}$. One of the most extensively used form of the free energy functional is the Ramakrishnan-Yussouff (RY) \cite{ramakrishnan79} free energy functional $F[\rho]$:
\begin{equation}
\label{RY}
\beta F[\rho]=\int \d{\bf x}\rho({\bf x})\bigg[\ln\frac{\rho({\bf x})}{\rho_0}-1\bigg]-\frac{1}{2}\int \d{\bf x}\d{\bf x}'\delta\rho({\bf x})c(x-x')\delta\rho({\bf x}'),
\end{equation}
where $\delta \rho({\bf x})=\rho({\bf x})-\rho_0$, $\rho_0$ being the homogeneous background density, the first term is the ideal gas contribution and the second term is the contribution due to interaction. $c(x-x')$ is the direct pair correlation function that contains the information of the inter-atomic interactions.

In the supercooled regime, the velocity field is slow and we will neglect the convective nonlinearity as well as higher order terms in the momentum density. Thus we expand
density and momentum density as
\begin{eqnarray}
\rho({\bf x},t)&=&\rho_0+\delta\rho({\bf x},t), \nonumber \\
{\bf g}({\bf x},t)&=& (\rho_0+\delta\rho({\bf x},t)){\bf v}({\bf x},t) = \rho_0{\bf v}({\bf x},t).
\end{eqnarray}

Using these simplifications, Eq. (\ref{densityeq}) and (\ref{momentumeq}) with the pressure term being replaced by the force density become
\begin{subequations}
\begin{align}
\frac{\partial \delta \rho({\bf x},t)}{\partial t} &= -\rho_0\nabla \cdot {\bf v}({\bf x},t) \label{linden55} \\
\rho_0\frac{\partial {\bf v}}{\partial t}=-\rho({\bf x},t)\nabla \frac{\delta F}{\delta\rho}&+\eta\bigtriangledown^2 V+(\zeta+\eta/3)\nabla(\nabla\cdot{\bf v})+{\bf f}({\bf x},t).
\label{linmom55}
\end{align}
\end{subequations}
Now, we take the divergence of Eq. (\ref{linmom55}) and use it in Eq. (\ref{linden55}) to obtain the equation of motion for the density fluctuation alone as
\begin{equation}
\frac{\partial^2\delta\rho({\bf x},t)}{\partial t^2}=D_L\bigtriangledown^2\frac{\partial\delta\rho({\bf x},t)}{\partial t}+\nabla\cdot\left(\rho\nabla\frac{\delta F}{\delta\rho}\right)-\nabla\cdot{\bf f}({\bf x},t),
\end{equation}
where $D_L=(\zeta+4\eta/3)/\rho_0$. After space Fourier transforming,
\begin{equation}
\frac{\partial^2\delta\rho_{\bf k}(t)}{\partial t^2}=-D_Lk^2\frac{\partial\delta\rho_{\bf k}(t)}{\partial t}+\left[\nabla\cdot\left(\rho\nabla\frac{\delta F}{\delta\rho}\right)\right]_{\bf k}+i{\bf k}\cdot{\bf f}_{\bf k}(t).
\end{equation}
Using Eq. (\ref{RY}) for the free energy functional, we obtain
\begin{align}
\label{shearfhmct1}
\frac{\partial^2\delta\rho_{\bf k}(t)}{\partial t^2}&+D_Lk^2\frac{\partial\delta\rho_{\bf k}(t)}{\partial t}+\frac{k^2k_BT}{S_k}\delta\rho_{\bf k}(t) \nonumber \\
&=k_BT\frac{\bf k}{2}\cdot\int_{\bf q}\bigg[{\bf q}c_q+({\bf k}-{\bf q})c_{k-q}\bigg]\delta\rho_{\bf q}(t)\delta\rho_{{\bf k}-{\bf q}}(t) +i{\bf k}\cdot{\bf f}_{\bf k}(t),
\end{align}
where the wave vectors are at time $t$. As we have seen in the previous section, under shear, we will have advection of wave vector and ${\bf k}$ at time $t=0$ will couple to ${\bf k}(t)$ at time $t$. The force density, ${\bf \mathcal{F}_{\bf k}(t)}$ is given as,
\begin{equation}
\label{shearfhmct2}
{\mathcal{F}}_{\bf k}(t)=\frac{k_BT}{2}\int_{\bf q}\hat{k}\cdot({\bf q}c_{\bf q}+({\bf k}-{\bf q})c_{{\bf k}-{\bf q}})\delta\rho_{\bf q}(t)\delta\rho_{{\bf k}-{\bf q}}(t).
\end{equation}
This force density, quadratic in density fluctuation, will have large fluctuations near the glass transition. In spirit of the Langevin equation \cite{zwanzigbook}, we can divide 
this term in two parts, one producing the damping and the other part being the noise \cite{kawasaki03}. Linear response theory is applicable close to equilibrium and the 
``new noise'' and the damping coefficient must be related as follows
\begin{align}
\mathcal{F}_{\bf k}(t)&=-\int_0^t \d s\mathcal{M}_{\bf k}(t-s)\frac{\partial \rho_{\bf k}(s)}{\partial s}+\xi_{\bf k}(t)\nonumber\\
&\mathcal{M}_{\bf k}(t)=\frac{\langle\xi_{{\bf k}(t)}(t)\xi_{-{\bf k}}(0)\rangle}{k_BTV},
\end{align}
where in the second equation we have explicitly used the time dependence on $k(t)$ to clarify the fact that this wave vector is at time $t$ when it's $k$ at $t=0$. With this form of noise, we will obtain the equation of motion for the normalised coherent intermediate scattering function as
\begin{equation}
\label{shearmct3}
\ddot{\phi}_{{\bf k}}(t)+D_Lk^2\frac{\partial }{\partial t}\phi_{{\bf k}}(t)+\Omega_{{k}(t)}^{HD}\phi_{{\bf k}}(t)+\int_0^t\mathcal{M}_{{\bf k}}(t-t')\dot{\phi}_{{\bf k}}(t')dt'=0,
\end{equation}
\end{widetext}
with the frequency term given by
\begin{equation}
\Omega_{{k}(t)}^{HD}=\frac{k(t)^2 k_BT}{S_{k(t)}},
\end{equation}
and the memory kernel is obtained same as in Eq. (\ref{shearMCT2}) that was obtained in the previous approach. The evolution equations (\ref{shearMCT1}) and (\ref{shearmct3}) differ slightly as we started from two different starting equations, but in the large density limit they lead to same time evolutions with a small difference at very short time.

We started from the equations of motion for a normal fluid, in case of colloid, $D_Lk^2$ will be replaced by $\zeta$, the friction coefficient as in the previous section. 
The input structural quantities of the theory are that of a sheared fluid. However, under the assumption of isotropic shear \cite{fuchs03}, the distorted structure factor becomes
same as the undistorted one. Our theory and those in Ref. \cite{fuchs02,fuchs03,fuchs05,miyazaki02,miyazaki04} differ in minor details but they all become qualitatively same under the schematic assumption. First, let us ignore the second order time derivatives in Eqs. (\ref{shearMCT1}) and (\ref{shearmct3}) as it only effects the short time dynamics. Then, after taking the isotropic assumption, we can write down the schematic equation of motion for the correlation function as
\begin{equation}
 \dot{\phi}(t)+\Gamma\phi(t)+\int_0^t m(t-t')\dot{\phi}(t')\d t'=0,
\end{equation}
with $\Gamma$, related to $\Omega_k$, gives the initial decay and $m(t)=\mathcal{G}(\gdot t)\nu\phi(t)^2$ is the memory kernel where $\nu$ gives the interaction strength and $\mathcal{G}(\gdot t)$ is a function chosen such that it decays at long time. $\mathcal{G}$ results from the fact that shear reduces the strength of the memory as a function of time. A number of forms are possible for $\mathcal{G}$, $e^{-\gdot t}$ being one of them [see \cite{fuchs03} for more details on this].

Here we have arrived at the theory with a series of transparent assumptions and the theory is similar to those derived earlier \cite{fuchs02,fuchs03,fuchs05,miyazaki02}. Some of the earlier approaches \cite{fuchs02,chong09,suzuki13} use an extension of the projection operator formalism where certain key steps, like the factorization of the four-density into products of two-density terms \cite{reichman05,mayer06}, are not apparently clear and whether they are valid arbitrarily far away from equilibrium is not obvious. Although our approach doesn't say anything about these approaches, it is interesting that we also reach to the same theory through the use of LRT. It indirectly shows that linear response theory might be important for the applicability of such approximations. A clear demonstration of this will be important for better understanding of MCT, even for a bulk unsheared system.

\section{Discussion}
\label{discussion}
The goal of the present paper is to understand the various approximations involved in the derivation of mode-coupling theory for sheared steady states and their domain of applicability. Such a task will be important for better understanding of MCT in general, even for a bulk unsheared system. In this work we obtained the theory for sheared steady state through two different approaches, first starting with the microscopic equations of motion of individual particles and then through the fluctuating hydrodynamics. The advantages of both the approaches compared to others (for example the projection operator formalism \cite{reichman05,chong09,suzuki13} or the integration through transients \cite{fuchs05}) are the transparency of various approximations. In our derivation, we see that one needs to make a number of approximations which can be justified only close to equilibrium. For example, in the first approach, the trial function is justifiable only if there is local equilibrium and the memory kernel is obtained through the use of linear response theory. In the second approach, again, one needs to use linear response theory and FDR. Within MCT, the memory kernel plays the major role and within the schematic approximation (where one ignores the wave-vector dependence of the correlation functions) some of the existing theories \cite{miyazaki02,fuchs05,fuchs03} become equivalent to ours. As we discussed in the introduction, a colloidal glass is far away from its structural arrest compared to a molecular glass \cite{durian99} and one can justify the use of linear-response theory for such a system when the shear is small. But one needs to be careful in applying these theories in general for systems arbitrarily far from equilibrium. One interesting question will be how to correctly treat the various currents within MCT outside the colloidal domain. There exists different approaches \cite{chong09,suzuki13} to this problem. However, the relations between various theories are not clear at the moment and we believe the current work will help drawing comparisons between different approaches.

As a byproduct of our calculation, we have obtained a generalized Yvon-Born-Green (YBG) equation for the sheared steady state through two different approaches. We show that the YBG equation yields the distorted structure factor if one assumes the random phase approximation (RPA). Such expression for the distorted structure factor was also obtained through different approaches \cite{ronis84,indrani95}. 

It would be interesting to extend the calculation for colloidal systems under strong confinement \cite{israelachvili88,hu91,persson94,saroj06,demirel96,alsten88}. The viscosity of a confined system becomes quite large and there is a glass-like transition \cite{saroj11,demirel96}. Interesting phenomena are observed in simulations \cite{vezirov13,mackay14,papenkort14,siems15} and experiments \cite{cohen04,lin14} when such systems are subjected to shear and sheared-MCT extended for confinement should capture these findings. However, this is a task outside the scope of the present work.

It would be important to extend MCT for sheared steady states of glassy and granular systems applicable even far away from equilibrium. We can accomplish this following a similar approach as was taken for spin-glass systems \cite{berthier00}. MCT has recently been extended for aging systems under shear \cite{saroj12,saroj13} that goes to a steady state when the waiting time $t_w$ becomes of the order of inverse shear rate. Then, if we take $t_w\to\infty$ limit of the equations, the resulting theory will describe a sheared steady state. As we haven't used any FDR-like relations in this theory, it should be applicable even far away from equilibrium. However, the cost we must pay for not using FDR is that we need to write down the equations for both correlation and response functions. 

\begin{acknowledgements}
We would like to thank Sriram Ramaswamy, Thomas Voigtmann and Kunimasa Miyazaki for many useful discussions.
\end{acknowledgements}

\bibliography{ref_shear}

\end{document}